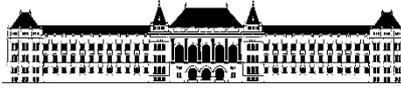

**Budapest University of Technology and Economics**
Faculty of Electrical Engineering and Informatics
Department of Computer Science and Information Theory

Mihály Péter Hanics

# GRAPH THEORETICAL MODELS AND ALGORITHMS OF PORTFOLIO COMPRESSION

ADVISORS:

DR. PÉTER BIRÓ, DR. TAMÁS FLEINER

BUDAPEST, 2022

# Contents

# Abstract

Matching problems under preferences gained significant attention not just in mathematical sciences, but in other fields too since the paper published by Gale and Shapley in 1962-ben on the college admission being modelled as a matching problem with preferences, and Al Roth's work on matching problems in practice, such as modelling kidney exchanges as matching markets. For their work on matching markets, Roth and Shapley received the 2012 Nobel Memorial Prize in Economics. Today, this interdisciplinary field extended to other parts of mathematics, for example, stability of flows is examined, and focuses on the real-life application of this field. One such application is financial clearing, or portfolio compression. Market participants (banks, organizations, companies, financial agents) sign contracts, creating liabilities between each other, which increases the systemic risk. Large, dense markets commonly can be compressed by reducing obligations without lowering the net notional of each participant (an example is if liabilities make a cycle between agents, then it is possible to reduce each of them altogether without any net notional changing), and our target is to eliminate as much excess notional as possible in practice (excess is defined as the difference between gross and net notional). A limiting factor which may reduce the effectiveness of the compression can be the preferences and priorities of compression participants, who may individually define conditions for the compression, which must be considered when designing the clearing process, otherwise a participant may bail out, resulting in the designed clearing process to be impossible to execute.

These markets can be well-represented with edge-weighted graphs. In this paper, I examine cases when preferences and conditions of participants on behalf of clearing are given, e.g., in what order would they pay back their liabilities (a key factor can be rate of interest) and I show a clearing algorithm for these problems. On top of that, since it is a common goal for the compression coordinating authority to maximize the compressed amount, I also show a method to compute the maximum volume conservative compression in a network. I further evaluate on the possibility of combining the two models. Examples and program code of the model are also shown, and pseudo-code of the clearing algorithms.

# 1 Introduction

This paper is a modified version of my submitted bachelor's thesis work at the Budapest University of Technology and Economics, at the Department of Computer Science and Information Theory, under the advisership of Dr. Tamás Fleiner (Department of Computer Science and Information Theory, BME) and Dr. Péter Biró (Institute of Economics, Centre for Economic and Regional Studies, Eötvös Lóránt Research Network). The paper describes the existing graph theoretical models of portfolio compression and expresses novel algorithms in the topic.

## 1.1 Financial clearing and its application

Trades, lending, and in general the exchange of goods have existed in our society since thousands of years [1]. People of different professions exchanged their product to other types of products from local inhabitants pursuing another profession. This method of exchange is bartering. Farmers traded their food gathered from their land, fishermen could offer seafood, a wool-man offered clothing made from the sheep's wool, other residents also took part in trades. Alternatively, they could sell their product for coins or any other medium (this method is believed by historians to have developed after bartering, see [1]). The spreading of currency eventually led to money lending, in Ancient Greece pawnbrokers already existed and lent money by collecting collateral from borrowers, which is common today as well. In Mesopotamia, Hammurabi defined the price of silver and regulated silver loans. Loans are very prominent even today, and are not necessarily centralized, but often are organized bilaterally, between two companies, or a bank and a citizen, or many other cases. Multilateral loans also exist but are rare in practice. Eventually, obligations are formed between lenders and borrowers, and these lenders and borrowers or investors and investees form a large network (named market) revolved around obligations, in which they're called market participants (otherwise financial agents, in the case of banks and companies). Theoretically, it can happen that two market participants are both lenders and borrowers to each other.

These markets regularly have room for compressing depths. This means, that the amount of some loans may be reduced, without any participant gaining any monetary benefit from it (nor any participant be disadvantaged). The action of clearing these debts collectively is portfolio compression. Financial clearing is a broader concept, it not only describes compression, but monetary allocation too, and redesigning the market (for example, when a company goes bankrupt), but in this paper we refer to it as reducing the obligations. The procedure of clearing is usually done in multiple cycles, such a clearing among a cycle can be seen on Figure 1-1, where each blue node represents a participant, and edges represent the amount of obligation from the buyer to the seller.

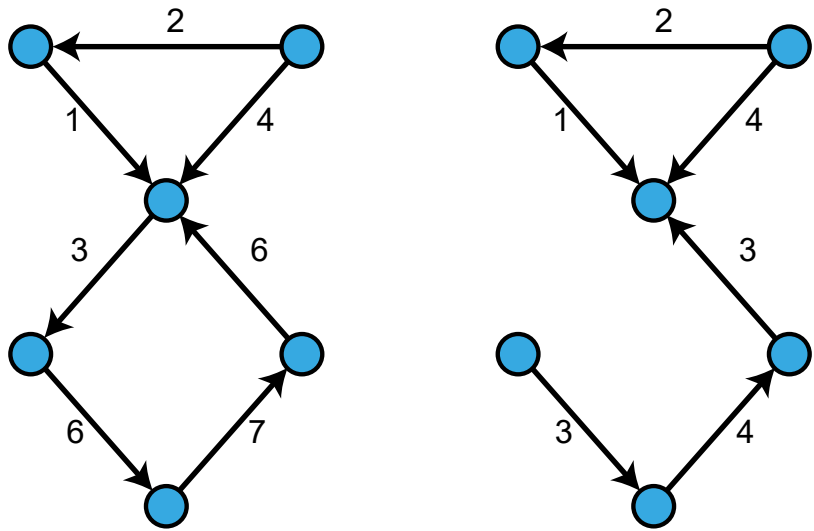

**Figure 1-1.: An example for clearing among a cycle, by 3 units.**

Networks of companies, organizations, banks, in the sense of financial obligations, form a market, that may be compressed. In practice, a provider called TriOptima sells its service to compress networks. Portfolio compression can be modelled using graph theory, with directed graphs.

## 1.2 Graphs, graph theory

Graph theory is a part mathematics for modelling networks, structures, frequently in other sciences too, such as economics, computer science, natural and social sciences, and electrical engineering (also in other field of mathematics, for example, operations research). Graphs are built up of vertices (also called nodes) and edges. Vertices may represent objects or in our context, participants, and edges between two vertices describe the relation between the participants the two vertices represent. Edges may have an orientation (the possible orientations are given by the vertex it ends in, and since it connects two vertices, there are two possible orientations for each edge), these are named directed edges (or simply arcs) and graphs containing directed edges are directed graphs. A walk is a sequence of edges where the node in which an edge ends is the start node for the next edge in the sequence. A walk is a circuit, if the last edge of the sequence ends in the node that the first edge comes from. A circuit is a cycle, if each node is the start and the end of edges in the sequence exactly once. A walk is a path, if each node is the start and the end of edges in the sequence exactly once, except for the first and last node of the sequence. Many laws and theorems from other sciences may be worded using the tools of graph theory. A great example of that is wording the most famous laws of circuit theory and electrical engineering: Kirchhoff's current law and Kirchhoff's voltage law. Kirchhoff defines nodes in an electric circuit's schematic, and states that the sum of current flowing into any node is equal to the sum of current flowing out of the node, alternatively, defining negative currents using directions, $\forall v_i, \sum_{j \mid \exists I_j} I_{ij} = 0$, where $v_i$ is a node, and $I_{ij}$ is defined if $v_i$ and $v_j$ are neighbours in the graph abstraction of the circuit schematic (thus, if there is at least one wire between them that contains exactly one electrical component between). On Figure 1-2, for example, nodes are represented as letters, these can be defined on the wires connecting the components, and wire segments that meet represent exactly one node. For example, the node labelled with letter C is the node represented by wires segments connecting three components: two impedances (dark grey rectangles), and a voltage source. Using Kirchhoff's current law on node C, assuming that positive current flows from nodes B and E to C, and from node C to D (thus $I_1, I_2, I_3 \geq 0$), we can derive that $I_2 + I_3 = I_1$. Kirchhoff's voltage law states that taking a closed loop (a directed cycle in graph theory is a closed loop in electrical engineering terminology), the sum of voltages is zero. This makes it possible to define a value called potential to each node, and from the law we get that summing the voltage differences across the closed loop marked with letter J, the sum of them is 0, thus $V_{AD} + V_{DC} + V_{CB} + V_{BA} = 0$. (Potential is also defined in vector fields, vector fields for which you can define a potential function are called conservative. For conservative fields, path-independence holds, and the electric field is conservative, thus we can say: on a path between two potentials, integrating the voltage difference along the path will always result in the difference of the two potentials.)

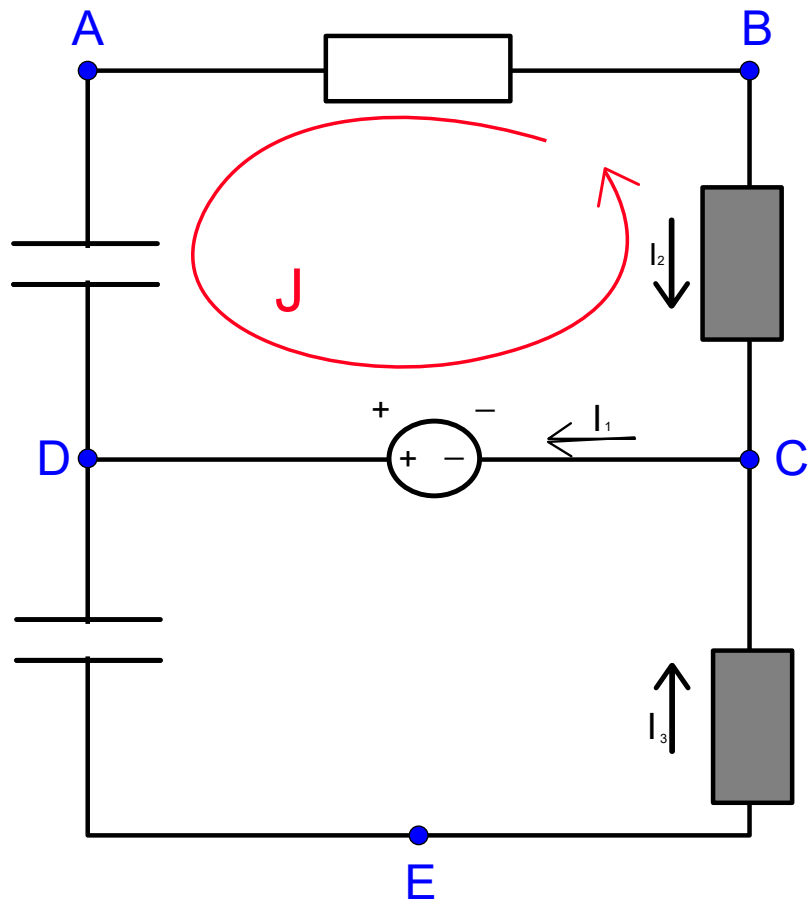

**Figure 1-2.: An electric circuit's schematic we use to describe Kirchhoff's laws.**

Determining circuit properties (values of voltages and currents mainly) is important and two common approaches (nodal analysis, also known as branch current method, and mesh analysis) build on these laws and handle the circuit as a directed graph.

Flow and circulation are two important terminologies in graph theory that were created for problems, where Kirchhoff's laws can be found, and they build on Kirchhoff's work. (Interestingly, this work is truly a precursor of graph theory, Kirchhoff work which not only inspired the idea of circulation, but his work defines junctions which are the equivalent of nodes in graph theory, publishing his work in 1845 whilst the first graph theory book was written in 1936 by Dénes Kőnig. Flows in a network were first studied in 1939 by Kantorovich [2].)

## 1.3 Application of graphs in operations research

Other research fields apply graph theory directly, not just taking inspiration from it. Sciences like economics makes great use of graphs, and graphs are often used for solving a practical challenge or modelling a real-world network. Euler used it on the "Seven Bridges of Königsberg" problem in 1736 to show that there is no way of passing on each bridge exactly once (without swimming across the Pregel River). In the early $20^{th}$ century, when graph theory was formulated and developed rapidly, it was foreseen how broadly can it be used. Newer and newer type of problems were formulated, such as matchings, colouring edges, and vertices, finding out whether a graph can be drawn on a paper without two edges crossing each other, and so on. Research on this found real-world usage, for example, matching problems are used today in kidney exchange, to find donor-patient pairs. Gale and Shapley helped graph theory be more common for practical use, they developed in some form optimal algorithms for problems like college admission, house allocation, and formulated matching under preferences, and stable matching. More recently, for example, computational social choice is one such field that is quite young, which inherited knowledge on the model of preferences. This paper also makes use of this model. In 2020, Milgrom and Wilson received the Nobel Memorial Prize in Economics for their work on matching markets (and auction theory). We can say that graphs are now used very frequently for modelling operations, and graph algorithms are used to find optimal algorithms for real-life tasks. The website http://www.matchu.ai/ visualizes some of the most common graph theory algorithms used for matching.

# 2 Current models of portfolio clearing and financial markets

There have been a substantial number of different models and approaches on portfolio compression, with most models aiming to reduce the gross notional in different markets, analysing these models from different point of views and more generally, quite a vast number of studies on financial clearing, describing the properties, benefits, and drawbacks of clearing models. These studies branch into three main fields: systemic risk, graph theory and freshly, game theory. (Other approaches like network analysis approaches were also published on this topic, see 2.5.) Systemic risk papers seem to be most common, however, for this thesis work, the papers describing the topic with graph theoretical approaches are most relevant. These types of markets can be distributed into two sets: over-the-counter (OTC) markets (decentralized) and centralized markets. Decentralized markets can be artificially centralized too, 2.4 discusses the act of adding a central coordinating party (CCP) to an over-the-counter market.

## 2.1 Graph theoretical approaches

The network of market participants and their obligations can be modelled by a graph $G = (V, A)$, where $V$ is the set of vertices (representing the participant) and $A$ is the set of capacitated arcs (weighted directed edges, representing the obligations). Clearing among a cycle in graph theory is taking a cycle in a graph and reducing each arc capacity by equal amount. (One may clear among a circuit as well.)

D'Errico and Roukny [3] formulated the graph theoretical description of the compression problem in 2017, and later republished their work in 2021. They described the network as a graph $G = (N, E)$, compressed into graph $G' = (N, E')$, with $N$ being the set of participants $\{1,2,3,…,n\}$, and $E$ and $E'$ being the set of gross exposures via contracts prior and after compression, respectively. They use a matrix structure for $E$, with size $n \times n$, where $e_{ij}$: i is the seller, j is the buyer, let $e_{ii} = 0$. (This structure can also be seen in Chapter 4.1.1.) Gross position and net position is defined for every market participant: $v_i^{gross} = \sum_j e_{ij} + \sum_j e_{ji} = \sum_j(e_{ij} + e_{ji})$, and $v_i^{net} =$ $\sum_j e_{ij} - \sum_j e_{ji} = \sum_j (e_{ij} - e_{ji})$). The goal then is to compress the graph into another graph where every notional (obligation) is minimised, under given conditions, thus the gross value shall be reduced as much as possible. They defined the term *excess* as the maximum amount of notional eligible for compression. Usually, only a fraction of the excess can be compressed, as it's limited by the conditions, which are regulatory constraints and portfolio preferences. Customers of the market are participants who only sell or only buy contracts, and dealers are the opposite: they sell at least one and buy at least one contract, they intermediate between other participants. We will use these in our models too. On compression, they note that in its minimalist form it's analogous to min-cost flow problem and defined tolerances on compression (compression tolerances are lower and upper bounds on each obligation, chosen such that the original market has to satisfy these tolerances, and the compressed network's obligations also have to satisfy these tolerances). Based on the preference settings, they describe two main compression types: conservative and nonconservative. Conservative compression only enables reducing the gross market value (and the amount of excess) on established obligations, thus among (already existing) edges in the graph representing the network. Nonconservative compression lets new contracts be made and does not limit the obligation values in any way.

    We can prove that a clearing of the network that reduces excess can be done via walks (with length at least 2) in the graph in the case of nonconservative compression, creating new directed edges between the two endpoints of the walks. It can be also shown that any conservative clearing can be broken down to only clearing among circuits (at least in the case of integer obligation values, a proof of this is in Chapter 3.2). From this, we can only compress nonconservatively if the market has at least one dealer and we can only compress conservatively if the graph of the market contains at least one directed cycle.

    Apart from conservative and nonconservative settings, the paper presents hybrid and bilateral settings. The latter only allowing clearing in cycles of length 2, and hybrid combining the first two settings, in which customer obligations have conservative constraints and dealer obligations have nonconservative constraints. The authors proved that they always rank in the same order in terms of efficiency.

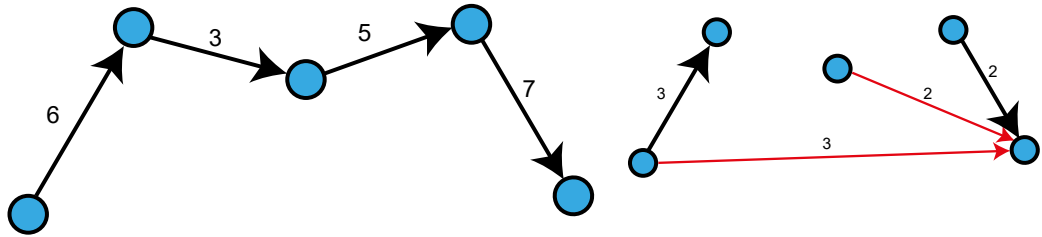

**Figure 2-1.: A possible outcome of nonconservative compression (original network on left, compressed on right).**

Amini and Feinstein [4] built on D'Errico and Roukny's work and showed some properties of the compression models and formulates the optimal network compression problem. On their generalization, they managed to prove that this problem, generically, is NP-hard for conservative, nonconservative and rerouting (no excess removed, liabilities are just redistributed) compressions.

In our paper, we will focus on the D'Errico and Roukny conservative model, our aim is to maximize cleared excess (thus our objective function is just a sum of all obligations, and we aim to minimalize it).

## 2.2 Systemic risk papers

First, we must talk about systemic risk in general. Systemic risk is a pointer to how a small event can cause industrial, geographical, economical, regional collapse, heavily damaging the economy. Networks that are heavily sensitive on changes in stability are more likely to experience systemic risk. Eisenberg and Noe [5] introduced systemic risk in financial networks. They studied that after a shock, how should obligations of all firms be determined based on the debt claims. They use a proportional model and show that there always exists a unique clearing payment vector. Rogers and Veraart [6] further developed Eisenberg's and Noe's model, and defined so called default costs: if a bank fails, they still have some default costs, and studied how this should be allocated among other banks, and showed that in cases, this may save a bank from going bankrupt after being affected by a shock. Compared to Eisenberg's and Noe's model, this rescue consortium might exist, and they stated some remarks on constructing one if it exists, whereas their model which doesn't include default losses has no incentive for solvent banks to rescue failing banks.

O'Kane [7] contributed more significantly firstly to the theoretical description of portfolio compression. In this paper, he shows that if the compression is designed optimally, counterparty risk should reduce after compression. Veraart [8] also studied portfolio compression, in terms of systemic risk, and showed that without defaults among firms, systemic risk is always reduced by compression, but this is not true otherwise.

## 2.3 Game theoretical models

Game theory is a field of mathematics that is commonly applied in other sciences, such as economics and social sciences. It studies strategical decision making in "games". Any game consists of three components: players (also known as agents), strategies and payoffs (goods). Given rules of the game, players make up strategies to acquire some payoff, while also considering strategies of other players (this concept is called *strategic interdependence*). Different generic strategies were formulated, and more recently, game rules were studied, about their properties. Three very important properties are: *individually rational, strategy-proof,* and *Pareto-efficient*. According to [9], a rule is individually rational if "at each problem it assigns an acceptable bundle to each agent", a rule is strategy-proof if "at each problem no agent can obtain a more preferred bundle by misrepresenting her preferences", and a rule is Pareto-efficient if "at each problem it is not possible to make all agents weakly better off and at least one agent strictly better off". Biró, Klijn, Pápai in [9] presented the properties of the circulation Top Trading Cycle (cTTC) rule, the Segmented Trading Cycle (STC) rules and serial rules in Shapley-Scarf exchange markets, where trades of goods are done in circulations (that is, exchanges are balanced, every agents gets as much in trade as he gave). For us, it's relevant that they showed that the cTTC rule is individually rational, but not Pareto-efficient or strategy-proof. Also, in [10] they formulated the cTTC and STC rules, and they showed that under circulations, there cannot exist a rule that is individually rational, strategy-proof, and Pareto-efficient.

Schuldenzucker, Seuken [11] considered debt contracts in portfolio compressing financial networks (ahead of default costs) and studied arising two questions: When is compression socially beneficial in a Pareto or a utilitarian welfare sense? Under which conditions do banks have an incentive to accept a given compression proposal? They showed that compression indeed be detrimental too, and gave sufficient conditions for banks to never come off worse after compressing.

Mayo and Wellman [12] studied the act of probabilistic shocks on external assets in portfolio compression. They gave a framework to approximate Nash equilibria if banks act accordingly to certain strategies.

## 2.4 Central coordinating models and decentralized clearing

In practice, it's common that an authority coordinates the compression itself. Commonly, when studying portfolio compression, we assume that clearing is settled between pairs of participants. However, it may happen that clearing is intermediated by the authority. This makes new methods for compression, centralized clearing with the assistance of a central counterparty (CCP).

O'Kean [7] illustrates central coordinated compression as seen on Figure 2-2. All obligations can be redirected to include the CCP as a middleman, then simple bilateral compression between the CCP and each participant. This leads to a very effective compression (not as effective as nonconservative compression), but in reality, participants do not want to lose their contact to other participants.

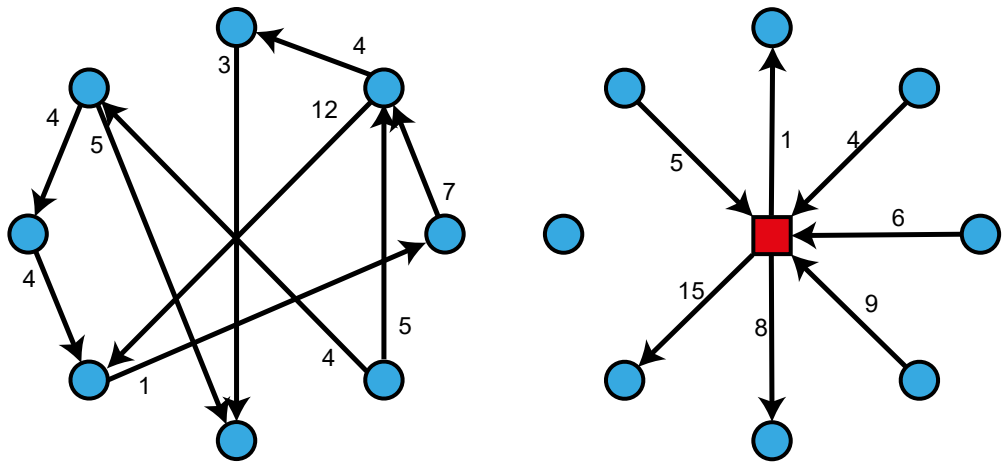

**Figure 2-2.: An example of central clearing, the red square representing the central clearing party. Original network on the left, compressed on the right**

Csóka and Herings [13] [14] [15] published multiple papers on the topic of centralized and decentralized clearing, and the differences between them. They introduced a class of (decentralized) clearing processes, where information like the payment matrices and equity values are not given, but the process in finitely many iterations converges to the least clearing payment matrix. They also studied centralized mechanisms, and gave linear programming formulations to them.

D'Errico and Roukny [3] stated in their paper that in their model, the from their empirical results, effectiveness of clearing when controlled centrally is less than when not centrally controlled.

## 2.5 Other papers

Other papers have also studied the underlying network structure behind these markets, often by the tools of game theory. Hatfield, Kominers, Nichifor, Ostrovsky, Westkamp [16] have given a model for bilateral contract trading, studying preference stability, and showed that with this model (under common circumstances), one can show that a stable output exists. Fleiner, Jagadeesan, Jankó and Teytelboym [17] also developed a model, in which, they designed trading networks accounting transaction taxes and commissions, and showed that if contracts have the properties of firms, then competitive equilibria exist, and the outcomes satisfy trail stability. Both models examine the network with complex preferences. Some papers show experimental results, empirical studies, but we will not focus on how well these models do in practice.

## 2.6 Notations

We conclude the common notations in this topic here.

-Financial networks:

A financial network is given by the triplet $(N, z, L)$, where $N = \{1, ..., n\}$ is the set of agents. Each agent starts with endowments, these are given by the vector $z \in \mathbb{R}_{++}^N$, $z_i$ represents all endowments of agent $i$. (Models below do not include values of initial endowments, $\forall i, z_i = 0$ ). $L_{++}^N$ is the liability matrix, $L_{ij}$ equals the amount of the liability agent $i$ has towards $j$, by convention, $L_{ii} = 0$, and let $L_i$ represent the set of liabilities of agent $i$. The letter $P$ most commonly represents the payment matrix, however we will not use that in our models. Here, it represents preferences. The preferences of agent $i$ is given by $P_i$, an ordered set. In the case of endowment preferences, $P_i$ is an ordered set with elements same as $N$, and in the case of obligation preferences, $P_i$ is an ordered set of obligations of agent $i$. The symbol $\prec$ means „more preferred", so $A \prec B$ means $B$ is more preferred than $A$.

-Graph notations:

Graphs are given by vertices and edges in-between vertices. A graph is most commonly labelled with letter $G$, if the graph is undirected, $G = (V, E)$ where $V = \{1, \ldots, n\}$ is the set of vertices, in our case, the set of market participants, and $E$ is the set containing the edges. If $G$ is a directed graph, then $G = (V, A)$, and $A$ is the set of arcs (directed edges). In the models below, we use this notation and represent networks as a directed graph. Let $A_{ij}$ equal the amount of obligation participant $i$ has towards participant $j$. We define functions on edges for the model described in Chapter 3.2: a capacity function $f: A \to \mathbb{R}$, a demand function $g: A \to \mathbb{R}$, a cost function $k: A \to \mathbb{R}$, and a circulation function $x: A \to \mathbb{R}$. For any $v \in V$, let $\delta_A^{in}(v) := \delta^{in}(v)$ be the set of arcs in $A$ entering $v$, and let $\delta_A^{out}(v) := \delta^{out}(v)$ be the set of arcs in $A$ leaving $v$, moreover, we let $\delta_A(v) = \delta(v)$ be the set of arcs. For any arc $(u, v) \in A$, $v$ is an outneighbour of $u$, and $u$ is an inneighbour of $v$ (as [2] says). The set of inneighbours of $v$ in $A$ is $N_A^{in}(v) := N^{in}(v)$, and the set of outneighbours of $v$ in $A$ is $N_A^{out}(v) := N^{out}(v)$. A function $x: A \to \mathbb{R}$ is called a circulation if for each vertex $v \in V$,

$$f\left(\delta^{in}(v)\right) = f(\delta^{out}(v))$$

that is the flow conservation law: the amount of flow entering a vertex is equal to the amount of flow leaving it. It is a similar concept to $s - t$ flows in graphs, an $s - t$ flow is a function for which the flow conservation law holds for each vertex except two vertices: one source node for which $\delta^{in}(v) = 0$, $\delta^{out}(v) > 0$, and one sink node, for which $\delta^{in}(v) > 0$, $\delta^{out}(v) = 0$.

As mentioned above, the main goal of compression is to reduce the gross value of the market, the more, the better. The fraction of excess cleared is the metric of how effective a compression is. Notional means the unit of obligations, liabilities.

(Common notation of circulations and flows is the letter *f*, and the upper and lower bounds are marked with letters *c* and *d* respectively, and another common notation is using letters *f* and *g*, and where *x* or *z* represents a flow or circulation, suggesting a connection to linear programming. We use the second notation.)

# 3 Novel approaches on portfolio compression with regard to agent preferences

Here I showcase my innovative approaches to this problem, where the agents' priorities are taken into consideration. The importance of satisfying agents' requests is clear, as if an agent is unsatisfied with the compression settings, the agent may deny the act of clearing.

The first model includes preferences as a priority: participants have a list of their obligations (or loans) in the order of how they would like to pay them back. The second model is an algorithm for conservatively clearing the maximum amount of excess. This rather satisfies the clearing coordinators, as their priority may be is to clear as much excess as possible, however this may be a useful approach for satisfying participants as well, as this gives the highest average for cleared excess per market participant, and participants may set a priority on at least how much excess from their obligations they'd like to clear.

These models describe the networks as graphs, as seen before, nodes represent the market participants, and arcs represent obligations. The orientation of the arcs is generally important, however, the algorithms presented below output the same result either way. For the sake of simplicity, we represent the obligation participant $i$ has towards $i$ with arc $a_{ij}$, from $i$ to $j$, (opposingly to D'Errico's and Roukny's model).

I build on D'Errico's and Roukny's model. The type of compression I use in these models are conservative (see 2.1, or [3]). This means, that the algorithms for each model only reduce among arcs that are already in the graph, thus these algorithms reduce obligations without establishing new ones.

The capacities of the arcs are the obligation amounts, and in these models are assumed to be integers. While it may happen in practice, that some obligations costs are described with a not integer number (say, if one unit equals one million dollars, then 1500000\$ equals 1.5 units), these models are not heavily affected by these, so we will neglect this occurring. Also, a thing to note, that obligation amounts converted into units will still remain rational numbers (two integers were divided), and any clearing algorithm that works assuming integer arc capacities can be modified to work for rational value arc capacities too. Just multiply all values with the least common multiple (LCM), run the algorithm, then evaluate the output of it knowing, that the original value of all arc capacities are the used values in the algorithm divided by the LCM.

There are two more important issues to take into consideration: the first one is that financial clearing can only be done on graphs that contain a directed cycle (else the conservative clearing algorithms do nothing as there is no cycle to clear), and the second one is that the graph describing a market may not be strongly connected or may be disconnected. (A graph *G* is strongly connected means that there is a node *A* from which there is no directed path leading to some other node *B*. Moreover, if there is no directed path from *B* to *A* either, then *G* is disconnected.) The first issue can be handled by previously checking whether the graph contains a directed cycle. To check whether a graph contains a directed cycle, a simple and conventional method is using the Depth-first search (DFS) algorithm, with a twist. The DFS algorithm traverses through a graph's nodes, by always choosing one of the nodes that are available and are farthest from the starting node (root node), only visiting each node once. Thus, it "always tries to go forward, until it can't, then it backtracks once and goes forward again till it can, then backtracks, and again, again, and again… until the algorithm stops". It labels each node with a number and those numbers represents the order of visiting a node (the root node is labelled with a 1, the last node is labelled with *|V|)*. If from a node, there is nowhere to traverse, as all of its neighbouring nodes are already visited, then the node's state is changed to "finished", otherwise, the node is "unfinished". The algorithm basically always tries to move forward from the (unfinished) node with the highest labelled number, else if it can't move forward, it changes the status of the node to unfinished, and backtracks once. Modifying this algorithm can help us find out whether there is a directed cycle in the graph: at each step when the algorithm picks a node to traverse from, check all of the nodes you can traverse to, and if there is any that's already visited, then the algorithm can exit with a "yes" (it contains a directed cycle), and if all nodes are already finished and we didn't exit yet, then exit with a "no". The algorithm exits with "no" if and only if the graph doesn't contain a directed cycle (such graph is called directed acyclic graph, or DAG for short). Then, we only try to compress when the network is not acyclic. (Note that on large graphs, even if they are quite sparse, there is a very high chance of a directed cycle existing. In practice, it's almost certain, so one may skip this step. The complexity of this algorithm is $O(|V| + |E|)$, which gives a relatively low runtime, but often there is no point in running it.) The second issue can be dealt with by checking how many components does the graph representing the network have and running the model algorithms separately on each component. Looking for strongly connected components is sufficient, as there cannot be a cycle in the graph that passes through nodes from more than one strongly connected component, else there would be a path in both directions from one component's nodes to the other component's nodes, which would mean that the two components are in fact one, which is contradiction. To find the strongly connected components of a directed graph, one may use Tarjan's algorithm [18] which is based on DFS and is linear time, or other algorithms (recently, *reachability-based* algorithms were studied and shown, but they are not better performance-wise).

## 3.1 Participants have preferences

In this model, we assume that market participants have preferences over which obligations they'd like to reduce. This is realistic, as each obligation differs, one may be more beneficial for one to clear than the other (for example: paying back manufacturers who produce a part of the product a company is selling may be more crucial). We can get inspiration to this model from matching under preferences, introduced by Gale and Shapley [19].

Matching under preferences is matching agents from different groups with each other, while matching with regard to the preferences of each agent. Gale and Shapley called such matching stable, if there is no two participants *A* and *B* such that they both prefer to be matched with each other than their current match. Later, other matching algorithms were developed, not necessarily focusing on stability (often from practical problems, like kidney exchange, finding algorithms for different matching markets). An algorithm for matching under preferences to take inspiration from is a trading algorithm, the Top Trading Cycle (TTC) algorithm, which is designed for the house allocation problem. It works this way: There are some people who each own an endowment (for example, a house), they are paired with their endowment. Everyone knows about each endowment, and they give a list of their preference of the endowments. The goal is to make trades such that everyone involved in a trade gets a preferred endowment (by their own standards). The algorithm takes each list and takes the first element (most preferred endowment) from each list and creates a graph where each vertex represents one person and his endowment and select arcs that will be put in the graph: for each person/endowment an arc from the it's vertex to the corresponding vertex of the endowment that is the first element of the person's preference list. Then the algorithm runs a node-deleting subroutine until the graph is completely empty. The subroutine: Generate the arcs in the graph, so make the set of arcs be the set of the first element of each preference list (the lists change after each subroutine loop). Find every cycle in the graph, and reallocate endowments among each cycle, such that each person receives the endowment that is first on their list. Next, remove everyone contained in these cycles and their endowments from the graph, and from every list. If the graph still contains at least one vertex, then run the subroutine again, else stop.

The algorithm stops when all agents/endowments were removed from the graph, and this will eventually happen. <u>Proof:</u> We need to show that we can always take another step (run the subroutine again) until the graph is empty. This can be proved by showing that the graph contains a cycle if it's not empty. If the graph currently contains $k$ number of nodes ($k > 0$), then the number of arcs in the graph is exactly $k$. Start from a random node and traverse along the leaving arcs. Since the number of arcs is greater than $|V| - 1 = k - 1$, at some point while traversing, we'll visit a node that was already visited previously. Say node $u$ is the vertex visited twice first. Then there is a cycle containing node $u$, formed by the path we took between visiting node $u$ the first time and the second time. (The walk we took between visiting the node twice is a path, else $u$ wouldn't be the node firstly visited twice.) Thus, a cycle exists, so the algorithm doesn't stop.

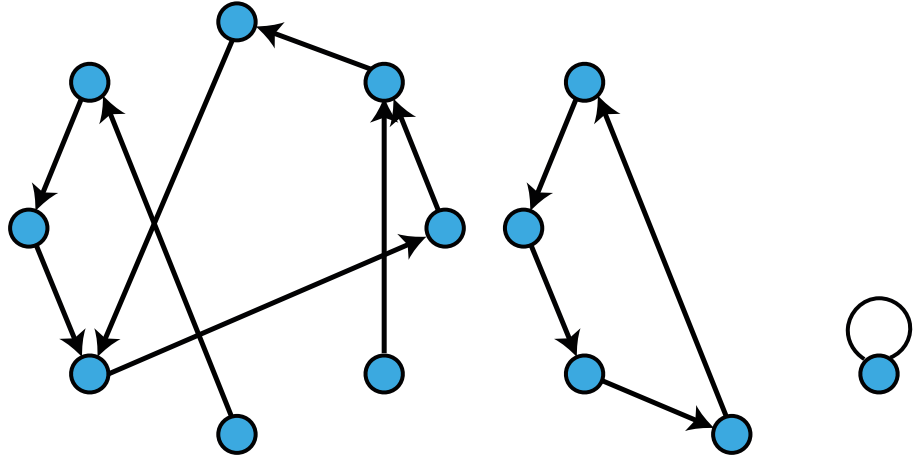

**Figure 3-1.: An example of the Top Trading Cycle algorithm, cycles are found in each step, then in the next step, agents of the cycles are taken out, and each agent chooses their most preferred endowment from the remainders. On the right is the result of taking out cycles in the first step**

This algorithm will result in each participant getting an "at least as preferred" endowment as their previous endowment, because the algorithm allows cycles of length 1, thus participants would pick their endowment over any less preferred ones. (This also means that the algorithm would do exactly the same if people didn't give a full list of their preferences, just the list of their preferences up until their own endowment.) For online interactive representation of the algorithm, check the MatchU website (mentioned in Chapter 1.3) and you can find a program for running the algorithm and under the House Allocation Problem menu item.

I keep the idea of generating a graph from a set of agents (from now on participants) and agent-associated ordered sets (preference lists). However, we need to slightly redesign the TTC algorithm for our problem, including the graph generation. We only run the algorithm on dealers of the market (as customers cannot be part of cycles). Firstly, the TTC algorithm uses an endowment-preference list. For us, that is not necessarily a better option than using an obligation-preference (alias arc-preference) list, because after clearing among arcs, it's more efficient to not remove a vertex that took part of clearing but instead just remove the arc between them, plus, it's more logical for us to store arcs with their current residual capacity in each loop of the algorithm (that is the difference between an arc's capacity and traversing flow). Also, in our model, we assume that multiple arcs from one node to the other do not appear, but these may theoretically exist, and node-preferences would not allow that. Let's define the graph generating routine: Given a set of vertices $V_s \subseteq V$ representing participants, a set of arcs $A$, and for each participant $i \in V_s$, $P_i$: an ordered subset of arcs in $\delta(i)$ that represents the preference participant $i$ has on them, we construct a graph $G_P = (V_s, A_P)$ such that $A_P$ consists of the most preferred arc from each preference list, that is $A_P = \{\forall i\ a_i : i \in V, a_i \in P_i, \forall a'_i \in P_i\ a'_i \prec a_i\}$. We will call this the *most preferred neighbours* graph of network $G = (V_s, A)$ with network preference $P = \{\forall i \in V_s\ P_i\}$. Moreover, if $P_i$ is an ordered subset of arcs in $\delta^{out}(i)$, this graph is the *most preferred outneighbours* graph, and if it is a subset of arcs in $\delta^{in}(i)$, then it is the *most preferred inneighbours* graph.

A simple algorithm for our problem iteratively works on the dynamically changing graph $G_{P^i} = (V_i, A_{P^i})$, starting from $i = 0, V_0 = V$, and $A_{P^0}$ is the set of arcs that are included twice in preference lists, that is, any obligation represented by arc $\forall a = (u,v) \in A$ that is in both $P_u$ and $P_v$. We assume that each arc has a capacity that is integer, so a function $f: A \to \mathbb{N}^0$ is given such that $f(a)$ is the capacity of arc $a$. The algorithm iterates until $V_i$ is an empty set. In each iteration, firstly, construct the most preferred outneighbours graph $G_i^C$ of network $G_{P^i}$ with preferences $P^i$, also make a copy of $G_{P^i}$ that is labelled with $G'_{P^i}$, then find all cycles in $G_i^C$. Let the cycles be $C^i = \{C_1^i, .. C_c^i\}$, and let $a_j^i = a: \min_{a \in C_j^i} f(a)$ be the arc with minimum capacity in cycle $C_j^i$. Of course, any arc in $G_i^C$ is also in $G'_{P^i}$. Now clear in $G'_{P^i}$ among each cycle with the maximum possible amount, so for any arc $a \in C_j^i \in G_i^C$, reduce the arc capacity to $f'(a) = f(a) - f(a_j^i)$ in $G'_{P^i}$. Whilst clearing, the algorithm removes any arc in $G_i^C$ with zero capacity and after clearing runs the node-removing subroutine. The node-removing subroutine removes nodes from $G_i^C$ that only have arcs going into them or arcs leaving them (or none) in $G'_{P^i}$, so if $v$ is a node in $G'_{P^i}$ for which $\delta(v) = \delta^{in}(v)$ or $\delta(v) = \delta^{out}(v)$, then we remove node $v$ and all arcs in $\delta(v)$ from $G'_{P^i}$, and if any nodes were removed, runs the node-removing subroutine again, else exits. In the last stage of the subroutine, increase the value of $i$ by 1 and let $G_{P^{i+1}} = G'_{P^i}$, and iterate if $V_{i+1}$ is not empty, then return. Eventually, the algorithm will terminate, this is because for any most preferred outneighbours graph, the number of arcs equals the number of vertices, hence a cycle must be contained, so in each iteration, we remove at least one arc from the graph, eventually emptying the set of arcs in the graph and hence the set of vertices too.

We need to prove however that an arc cannot appear in two cycles in any most preferred outneighbours graph, so surely no violations occur while clearing (else it may happen that the common arc in two cycles will have negative capacity after clearing), and this would show that the order of clearing cycles does not matter. This can be shown using the fact that in the graph, any node has exactly one outneighbour, from the construction of the graph. Indirectly, assume that the graph contains two cycles $C_A$ and $C_B$, which contain at least one common arc. Then an arc $a = (u, v)$ contained in both cycles exists such that $(v, v_A) \in C_A$, $(v, v_B) \in C_B$, and $v_A \neq v_B$. This would mean that node $v$ has (at least) two outneighbours: $v_A, v_B$, which contradicts our construction. Therefore, it is not possible that two cycles share a common arc. Moreover, this shows that $C$ is a set of disjoint cycles.

This concludes that the algorithm in each iteration rightfully clears among at least one cycle, and it will terminate, in at most $m$ iterations.

## 3.1.1 Obligation-ordering $\frac{1}{\varepsilon}$-preference algorithm

The above algorithm needs modifying if it's not a necessity or goal to clear totally among arcs. We can easily generalize the above algorithm by letting arcs be removed from the lists (assume them to be cleared) if the amount of flow passing reaches at least $x\%$ of their capacity, with $x$ being given. Thus, we define a number $\frac{1}{\varepsilon}$ for a bound (where $0 < \varepsilon \leq 1$): if an arc $a$ has at least $\varepsilon f(a)$ flow traversing through it, then we can mark the arc as "finished", and each time at 3) we remove these finished arcs from the lists too. This gives different results than just choosing $\varepsilon f(a)$ as the capacity of the arc in the algorithm, because with this way, a "cleared" arc has had $\varepsilon f(a) \leq x(a) \leq f(a)$ flow traversing through it, whilst the other way we will at most have $\varepsilon f(a)$ flow traversing through the arc. (It doesn't make sense for $\varepsilon$ to be 0, because that would mean that we would remove all arcs in the first iteration as 0 flow is enough for arc removal, hence why I note $\frac{1}{\varepsilon}$, as a reminder that $\varepsilon$ cannot be zero. If it is needed to remove an arc after reducing its capacity, then let $\varepsilon$ be a very small number greater than 0, one can show that such small number exists such that using this algorithm, any clearance on arc $a$ will result in more clearance than $\varepsilon f(a)$.)

If we further generalize this with letting $\varepsilon$ be a function of arcs, then we remove an arc from the preference lists if the flow passing through the arc is at least $\varepsilon(a)k(a)$, which is in fact, equivalent to having lower bounds for each arc.

### 3.1.2 Participants rank their loans, instead of obligations in which order they would like to settle them

We can use the same algorithm if we have the preferences of obligations that participants are waiting to be paid to them. In this case, we construct the most preferred inneighbours graph in each iteration instead. However, I do not believe this is more applicable in practice, it seems more forced due to participants picking the participant to pay their loans back, likely it results in a network with more systemic risk.

### 3.1.3 Further notes and an example

The node-removing algorithm is essentially removing nodes that are not strongly connected to any other node in the graph. It shares similarities to dividing the graph into strongly connected components, such as Tarjan's algorithm, which we mentioned earlier.

Biró, Klijn, and Pápai [9] showed that any circulational exchange (this is one such exchange, as the linear combination of cycles result in a circulation) is not strategy-proof, meaning that if participants do not give their true preferences, they might gain more benefit, or in the case of debt contracts, manipulate the market to gain more notional if a participant goes bankrupt. Therefore, preferences are needed to be dealt with carefully in practice.

At last, let's show the algorithm on an example graph, presented on the left side of Figure 3-2:

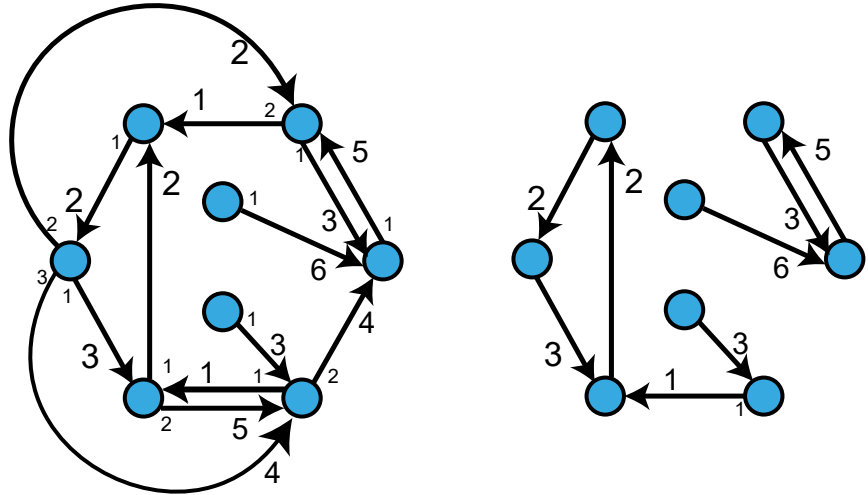

**Figure 3-2.: An example graph on the left and the most preferred outneighbours graph constructed from it.**

The graph on the left $G_{P^0} = (V, A_{P^0})$ represents 8 participants, the bigger in size numbers at the end of arcs represent the capacity, and the smaller size numbers at the start of arcs represent priority order. First, the algorithm creates the most preferred outneighbours graph $G_0^C$, containing the participants, and arcs with highest priority (marked with a 1). The graph on the right is the graph constructed, and simply by going along the arcs, the algorithm finds all cycles, in this example, two cycles were found. Then it reduces arc capacities among the two cycles in $G'_{P^0}$, the copy of the original graph, and removes all arcs with capacity the aftermath represented on the left side of Figure 3-3. Now it looks for nodes that only have arcs going into them or only leaving them and removes these nodes with their arcs from $G'_{P^0}$. The remaining graph can be seen in the middle of Figure 3-3.

Since the subroutine removed nodes and arcs, it runs the subroutine again, which removes node $A$ from $G'_{P^0}$, then runs the subroutine again, which removes node $B$ from $G'_{P^0}$. The subroutine runs again but doesn't remove another node, so it exits and the graph on the right which is the latest state of $G'_{P^0}$ will be $G_{P^1}$. The next iteration starts because the graph is not empty, and in this iteration, firstly finds the only cycle, clears among it, reduces both capacities by 1, removes the arc with now 0 capacity, and then removes the two nodes in the subroutine, which runs again, then exits, $i$ is increased to 2, but $G_{P^2} = G'_{P^1}$ is an empty graph, so no more iterations are done.

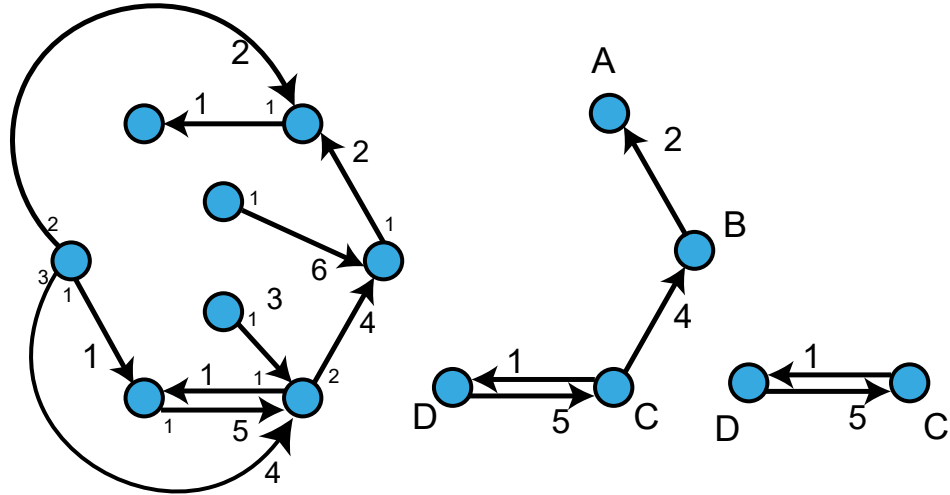

**Figure 3-3.: The result after an iteration.**

Overall, we cleared among 3 cycles, 7 arcs, 6 of which were highest priority arcs.

(The node removing subroutine could be improved to run less, if it directly looked at strongly connected components from the get-go (possibly on $G_{P^0}$).)

## 3.2 Conservative portfolio compression via circulations, maximizing the volume of clearable excess

While the models mentioned in Chapter 2 clear via finding cycles (except for 2.4), it is not necessary to try to clear by finding cycles. All we need to guarantee is that after the clearing, each participant's net notional doesn't change. This means that we just need to find a function of the arcs, such that adding the function values where the argument is an arc entering a node is equal to the sum of the function values where the argument of the function is an arc exiting the node. We then can reduce each capacity (obligation) with the value of the function applied on that arc. Such function exists in graph theory, it is called circulation. We defined what a circulation function is in 2.6. Furthermore, given a circulation $x$ and upper and lower bound $f$ and $g$, $x$ is a feasible circulation if $\forall a\ f(a) \geq x(a) \geq g(a)$.

The lower bounds are often called demands, and the upper bounds are called capacities. These are optimally not strict bounds, equality is allowed, this lets linear programming approaches be made (in this paper, we do not present any such approach, but there is room for future research).

The existence of a feasible circulation in a graph can be determined by Hoffman's circulation theorem (see [2]). Finding a feasible circulation (if it exists) can be done straightforward by transforming the circulation-finding problem into a max-flow finding problem, [2] elaborates more on this. In our model, we won't have any problem of a circulation existing in a given graph, because we well set the demand function $g$ to be 0. This guarantees that circulations exist in our graph (taking $x: A \to \mathbb{Z}$, $x(a) = 0$ for example, $x$ is a circulation in the graph).

We define the cost function on arcs: moving flow through an arc will have a cost, and if we define a cost function $k: A \to \mathbb{R}$ on the arcs, then the cost of moving $x$ amount of flow through arc $a$ equals $k(a)x(a)$. Flows and circulations have costs too, that is the sum of all costs in the graph, that is $\sum_{a \in A} k(a)x(a)$. Generally, there is a set of problems on minimizing cost, such as minimum cost circulation, minimum cost flow, minimum cost transhipment, and algorithmically finding these minimum functions in a graph. These are all related, furthermore, the minimum cost circulation and minimal cost flow problems are in fact equivalent, for a proof, see [20]. Unique, special cost functions lead to other known problems in graph theory. For example, if we set all costs to 1 in the minimum cost flow problem, then this is equivalent to the shortest path between two nodes problem. A maximum *s-t* flow problem also can be deduced from the minimal cost flow problem, just set the cost of arcs leaving $s$ to $-1$, and all other arcs costs to 0. Moreover, our approach will be deduced to a minimum cost circulation problem.

In this model, we clear excess among a circulation, and we'd like to select "the most optimal" circulation for clearing. The model defines "most optimal" as the circulation with maximum volume, that is; if we sum all the flow passing through an arc, from arc to arc, then pick the circulation where this sum is the biggest. In other words, we choose circulation $x$ for which $\sum_{a \in A} x(a)$ is maximal. It sounds counter-intuitive, but this is also equivalent to the minimal cost circulation problem, we just have to choose the cost function to be $-1$ everywhere. We're minimalizing a sum of negative terms, then multiplying it by $-1$, it's as if we were maximizing a sum of positive terms.

In our case, we use integer costs, and nonnegative integer capacities, so $k, f: A \to \mathbb{N}^0$, with demand $g = 0$, furthermore $k = -1$ (and $k(a) = 1$ on some arcs in the residual graph, more on that later). If the capacities, costs, and demands are integer, then one can show that there exists a circulation with integer values. Moreover, in this case, a circulation can be split to multiple cycles, such that adding up the flows (thus, the circulation is a unique linear combination of directed cycles). My simplest proof for this: Let's assume we have an integer circulation $x$ in graph $G = (V, A)$. Construct a graph $G' = (V, A')$ if $a \in A \leftrightarrow a \in A'$, and the capacity $f'(a)$ of an arc in $G'$ equals $x(a)$. This means that in $G'$, the capacity function is a circulation. Until there is no directed cycle in the graph, keep iterating: Choose a directed cycle in $G'$ and reduce all of the capacities in the cycle by the smallest capacity in the cycle, and remove any arcs if they now have 0 capacity. After any iteration, the capacity function will still be a circulation, and after finite number of iterations, there will be no cycles left in the graph. If the graph consists of no arcs, then we're done, that means that a circulation can be divided to cycles. Assume that there is at least one arc still in the graph, then start from one arc, and traverse forwards on the arcs till there is no way to traverse further. We cannot visit nodes twice, else there would be a cycle in the graph. If we stopped at node $v$, then there is no arc leaving $v$, but there is at least one coming into it, which would mean that $f'\left(\delta^{in}(v)\right) > 0$, $f'(\delta^{out}(v)) = 0$, but since after every iteration the capacity function still is a circulation, we get a contradiction. Thus, our assumption is correct.

We now can describe the algorithm we will use, to find a maximum volume circulation.

Algorithm (Goldberg-Tarjan '89):

1) $G = (V, A, f, g, k)$. Set $\forall a \in A$, $g(a) = 0$, and we set $\forall a \in A$, $k(a) = -1$ for all arcs (thus the function $g$ is 0 and $k$ is -1 everywhere).

2) Run a (strongly polynomial time) minimum-cost circulation algorithm on graph $G$, and find the minimum cost circulation for compressing.

3) Acquire the maximum volume circulation by the cost of the output (by the sum of each function value).

(Alternatively, on 2) we may check first whether the graph residual graph contains a negative cycle, and if it does, only then we look for a minimum mean cycle. More on this below.)

At 2), commonly, solutions for the minimum-cost circulation problem use the *negative cycle search* method. By itself, if we do not choose the cycles carefully, it may only result in only a quasi-polynomial time algorithm. According to [2], Weintraub in 1974 proposed two approaches that were shown afterwards to only result in exponential algorithms. I describe what is needed to know about the negative cycle search:

We define the residual graph for a graph $G = (V, A)$ and circulation $x$ in it. It shows how much more flow can be sent on each arc, and how much flow can be "reversed" (sent back). For any arc $a = (u, v) \in A: f(a) > x(a)$, define a "forward" arc in the residual graph $G_f = (V, A_f)$: $a' = (u, v)$ with capacity $f'(a') = f(a) - x(a)$, and for any arc $a \in A: x(a) > g(a)$, define a "reverse" arc in the opposite direction: $a'' = (v, u)$, with capacity $f'(a'') = x(a)$.

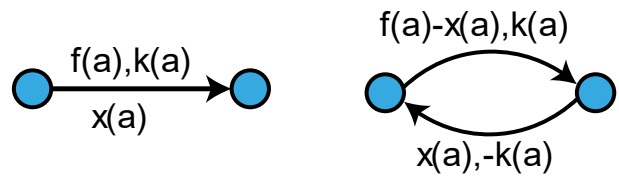

**Figure 3-4.: An arc from the original graph on the left, is the equivalent of two arcs in the residual graph**

Thus, the residual graph's set of arcs $A_f$ is defined this way: $A_f := \{a \mid a \in A, x(a) < f(a)\} \cup \{a^{-1} \mid a \in A, x(a) > g(a)\}$, where if $a = (u, v) \in A$, then $a^{-1} = (v, u)$. Examining this new graph lets us find augmenting cycles (similar to augmenting paths in the Ford-Fulkerson algorithm for finding maximum flow) which lead to improving circulation cost among these cycles. Define a cost function for these newly added arcs: $k(a') = -k(a'') = k(a)$. This way, if we can send σ amount of flow through a cycle $X^C$ (backwards flows are also allowed), then the difference in cost would equal to the difference in cost in the residual graph if we sent some flow among $X^C$ in the residual graph. ($X^C$ is also in the residual graph.) This suggests a method for minimalizing cost, which uses the following theorem:

<u>Theorem</u>: Let $G = (V, A)$ be a directed graph with a demand function $g$, a capacity function $f$, and a cost function $k$. Assume $x$ is a feasible circulation in $G$. Then $x$ is a minimal cost feasible circulation if and only if the residual graph of $G$ contains no directed circuit with negative cost.

(Note: for any circuit, here we can just sum the cost values of the arcs of the cycle and define that as the cost of the cycle. Because if we want to traverse σ more amount of flow through the cycle X, then the overall cost changes by σ · c(X), where c(X) is the cost of cycle X.) For a proof of the theorem, see [2]. This theorem is what motivates the "negative cycle search" method, which uses this fact to search for circulation improvements until there is no negative cost cycle in the residual graph.

To find a polynomial time algorithm for the problem, Edmonds and Karp invented a method called *scaling technique* (with which they actually gave the first polynomial time algorithm for the minimum cost flow problem in 1972). Many other algorithms developed later on had this implemented in its model. The technique can be implemented to run in $O(m(m + n \log n) \log C)$ time. The runtime becomes longer by choosing larger capacities, which means, that it's only a weakly polynomial algorithm.

<u>Definition:</u> A polynomial algorithm on integer input values is weakly polynomial if the complexity is dependent on the values of the inputs. For example, if the values are large, runtime may be considerably longer than for small values. A strongly polynomial algorithm is a polynomial algorithm where the complexity is not affected by values of the inputs.

Tardos [21] in 1985 managed to prove that a strongly polynomial algorithm exists. In 1988, proceeding the 20[th] annual ACM symposium, Goldberg and Tarjan showed a strongly polynomial algorithm, published in 1989 [22] proposing the *minimum mean cycle* method (see [23]). This method is considerably simpler than other methods at that time (although not as efficient, as of that time, but later was improved). We will use this in our model. Other fast algorithms use the network simplex method (which is often very fast, but it can at cases be exponential time), which is a linear programming approach.

Based on all stated above, we extend 1) with: "Set the circulation function $x_0 = 0$, and $i = 0$.". On top of that, we split 2) into two steps: <u>First step:</u> Create the residual graph $G_{x_i} = (V, A_{x_i})$ from graph $G$ and circulation $x_i$, and find a minimum mean cycle $X^C$ in $G_{x_i}$. Let $-\varepsilon_i$ be the value of the mean. <u>Second step:</u> If $-\varepsilon_i < 0$ then set $x_{i+1} = x_i + \tau X^C$, where $\tau$ is the maximal possible amount to increase or decrease each flow in $X^C$ (we add $\tau$ or take away $\tau$ from each arc in $X^C$, based on if flow is sent on a forward or reverse arc in the residual graph), we add 1 to $i$ and we go back to the first step in 2). Else, we know our graph is minimal cost, and we go to 3).

Karp showed that finding a minimum mean cost cycle can be done in $O(mn)$ time. This algorithm only works on strongly connected graphs, thus as suggested above, first separate the different components of the graph, and run this on each of them. His algorithm, published in 1972: Let $0 \leq k \leq n$, $d_k(v)$ be the minimum distance walk ending in vertex $v$ with $k$ arcs. We define this function as so: $\forall v\ d_0(v) = 0, d_{k+1}(v) = \min\{d_k(u) + c(a) | a = (u, v) \in A\}$, because the lowest distance $k+1$ arc walk to vertex $v$ is a lowest distance walk with $k$ arcs to some vertex u plus the distance between $u$ and $v$. Karp managed to prove that minimum mean of a directed cycle is equal to

$$\min_{u \in V} \max_{0 \leq k \leq n-1} \frac{d_n(u) - d_k(u)}{n - k}$$

Furthermore, a minimum mean cycle is contained in the path that set $d_n(u)$. We will use this to find $X^C$ in 2), the distance function is the cost function.

Note: There are different algorithms for finding the minimum mean cycle. See [24] for a comparison between the most fundamental approaches on how they do in real-world graphs.

This way of computing the value of $d_k(v)$ from the previous value $d_{k-1}(v)$ is a dynamic programming approach. Dynamic programming is a very strong tool, for some problems, the only known type of algorithms are

Goldberg and Tarjan showed that by always choosing the minimum mean cost cycle (that takes at most $O(mn)$ time to find) to reset the circulation function, we will receive the minimal cost circulation in at most $O(mn * nm^2 \ln(n))$ time (this is true for real numbers, for integers, it is even faster). This concludes that this algorithm runs in strongly polynomial time.

## 3.2.1 An example of the maximum volume compression algorithm working

Take the network represented by the graph on the left of Figure 3-5, and compute the maximum volume circulation for compression, by our method above. We describe each step of our algorithm on this network.

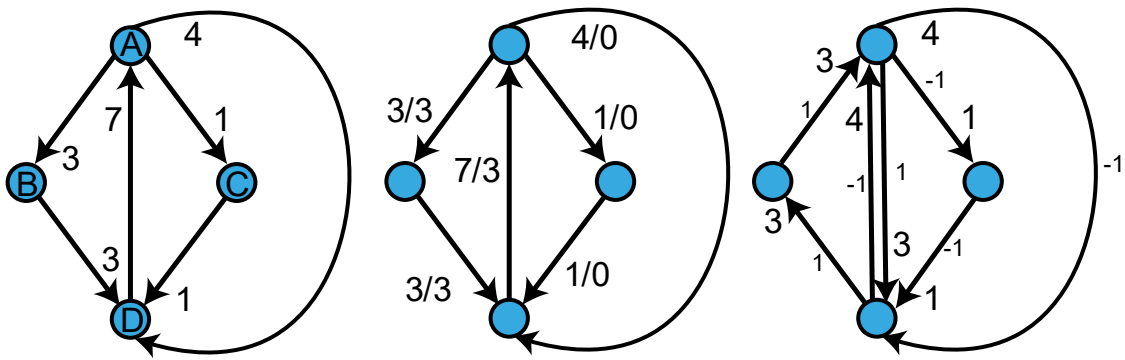

**Figure 3-5.: A network to compress on the left, the flow distribution after finding the first minimum mean cycle in the middle, and the residual graph on the right.**

Note that all costs in $G$ are -1, and we assume that the starting circulation $x = 0$. From the graph, we find that $n = 4$. The algorithm first constructs the residual graph, which is as of yet, equivalent to $G$.

From the residual graph, it computes $d_k(u)$ for all $u \in V, 0 \leq k \leq n$. It does it by filling the matrix below labelled with 1), where the $k + 1$th column from the left represents $d_k$, and the rows represent $A, B, C, D$, respectively (the matrix below it represents the arc that set $d_{k+1}(v)$). Since all arc costs now are $-1$, any 4-arc walk has $-4$ cost, and any cycle has $-1$ cost, from the equation above we also get that $-1$ is the minimum mean. In the arc table, the first three rows are fixed as there is no other arc to end the walk into vertex $A, B$, or $C$, the last row can be different, but assume it's the one noted. Any vertex could have set the minimum, but let's assume it was vertex $A$. Then the walk setting $d_4(A)$ contains a minimum mean cycle of mean $-1$, and from the table below we can reverse engineer the walk: the walk backwards is $AD - DB - BA - AD$, so the walk that contains a minimum mean cycle is $D - A - B - D$, from which we find that the cycle D-A-B, which has cost $-1$, so we found a minimum mean cost cycle.

$$1) \begin{bmatrix} 0 & -1 & -2 & -3 & -4 \\ 0 & -1 & -2 & -3 & -4 \\ 0 & -1 & -2 & -3 & -4 \\ 0 & -1 & -2 & -3 & -4 \end{bmatrix} \quad 2) \begin{bmatrix} 0 & -1 & -2 & -3 & -4 \\ 0 & 1 & 0 & -1 & -2 \\ 0 & -1 & -2 & -3 & -4 \\ 0 & -1 & -2 & -3 & -4 \end{bmatrix}$$

$$1) \begin{bmatrix} \emptyset & DA & DA & DA & DA \\ \emptyset & AB & AB & AB & AB \\ \emptyset & AC & AC & AC & AC \\ \emptyset & BD & AD & BD & CD \end{bmatrix} \quad 2) \begin{bmatrix} \emptyset & DA & DA & DA & DA \\ \emptyset & DB & DB & DB & DB \\ \emptyset & AC & AC & AC & AC \\ \emptyset & AD & AD & AD & CD \end{bmatrix}$$

We now send flow among the cycle, acquiring a new circulation $x_1$ within $G$, seen in the middle of Figure 3-5. The algorithm now computes the residual graph $G_{x_1}$, which is on the right of Figure 3-5. In $G_{x_1}$, it computes each least distance walk with $0 \leq k \leq n$ and stores these values and distance setting arcs in 2). Again, we have a minimum mean cycle of mean $-1$, say it was node $A$ that set the minimum in the equation above, we find that the $d_4(A)$ setting walk is $D - A - D - A$, and contains the cycle $D - A$ with minimum mean cost, again, we send flow on the cycle, and obtain the circulation $x_2$, seen on the left side of Figure 3-6.

Now again, the algorithm computes the residual graph $G_{x_2}$ from $G$ and $x_2$, seen on the middle of Figure 3-6. Compute the matrices below again. Computing the minimum mean cost, we find that $\max_{0 \leq k \leq 3} \frac{d_4(A) - d_k(A)}{4-k} = \max_{0 \leq k \leq 3} \frac{d_4(B) - d_k(B)}{4-k} = \frac{1}{3}$, $\max_{0 \leq k \leq 3} \frac{d_4(C) - d_k(C)}{4-k} = -\frac{1}{3}$, and , $\max_{0 \leq k \leq 3} \frac{d_4(D) - d_k(D)}{4-k} = 0$. The algorithm chooses the node for which the previous expression is minimal, and that is node $C$, for which the mean value is $-\frac{1}{3}$. The value is still negative, which means we still can improve the cost of our circulation. The algorithm then searches for the cycle in the path that set $d_4(C)$, which is C-A-D-C. This contains cycle $A - D - C$ with cost $1 + (-1) + (-1) = -1$, with three arcs, the with mean cost $-\frac{1}{3}$, as we wanted. We can send flow through this cycle too, the algorithm finds that it can send at most 1 unit of flow and finds the $x_3$ circulation to the one seen on the right side of Figure 3-6.

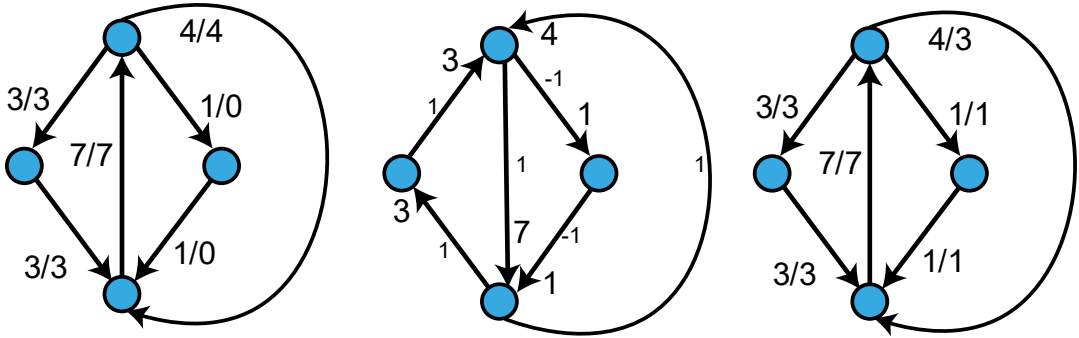

**Figure 3-6.: Circulations on the sides, and a residual graph in the middle, of the graph on the left.**

$$3) \begin{bmatrix} 0 & 1 & 0 & -1 & 0 \\ 0 & 1 & 0 & -1 & 0 \\ 0 & -1 & 0 & -1 & -2 \\ 0 & -1 & -2 & -1 & -2 \end{bmatrix} \quad 4) \begin{bmatrix} 0 & 1 & 0 & 1 & 0 \\ 0 & 1 & 2 & 1 & 2 \\ 0 & 1 & 2 & 1 & 2 \\ 0 & -1 & 0 & -1 & 0 \end{bmatrix}$$

$$3) \begin{bmatrix} \emptyset & DA & DA & DA & DA \\ \emptyset & AB & AB & AB & AB \\ \emptyset & AC & AC & AC & AC \\ \emptyset & CD & CD & CD & CD \end{bmatrix} \quad 4) \begin{bmatrix} \emptyset & DA & DA & DA & DA \\ \emptyset & AB & AB & AB & AB \\ \emptyset & AC & AC & AC & AC \\ \emptyset & AD & AD & AD & AD \end{bmatrix}$$

At last, we compute the residual graph $G_{x_3}$ and the corresponding matrices for the walks, and we find that all four vertices give nonnegative value in the equation above, the minimum mean cost is computed to be 0. This means that we can no longer improve our circulation, and the algorithm terminates, finding the maximum volume circulation, shown on the right side of Figure 3-6.

# 4 Programming the model and algorithms, with examples

Proving that our models are realistic is important. In this chapter, I present program code for evaluating the solutions of each approach. This is to show that it is indeed possible to implement these methods, and this code is rather illustrative, than being an actual (web-)app. (That shouldn't be our goal, as there is no universal data storage for OTC markets in practice, and the first thing the problem should do is acquire data from a data base, which it wouldn't be able to automatically, unless we know the way the data is stored). Before presenting the solutions, I describe the technical challenges and considerations raised.

The models shall be implemented in a programming language and environment that is flexible and commonly used for graph structures. Aside from that, it's important to create programs which can be easily adjusted or integrated into other programs or projects (if a professional is looking to use one of these models on a large dataset acquired from external sources, then possibly, the dataset is stored in a mischievous database, especially if it contains data from previous decades, when technology was more restricted, considering this we shall create a solution that is flexible for these cases). Three possibilities are mentioned here.

The first one is writing the code in the Python programming language and using the language's libraries used for discrete mathematics. Python is one of the most versatile programming languages existing (capable of web app development, scripting, functional programming, etc.), is arguably the second most popular programming language [25], focusing on easing the integration of public libraries into any code, to help developers use code written by someone before. Two of the most common graph theory libraries for Python are NetworkX and igraph (other options include modules like graph-tool, developed by CEU network scientist and physicist Tiago Peixoto which is newer and much faster (see [Performance Comparison - graph-tool: Efficent network analysis with python (skewed.de)](#)), and even contains methods developed while researching, with great visualization tools). The community contributes heavily, and there are good libraries for Python where graphs are already implemented as a data structure and common functions, algorithms. Python is not a fast language however, which shall be taken into consideration, when using large data sets, where an algorithm may run for hours or even days. NetworkX deals with large data sets well (although users online have stated that igraph is way more efficient). A well-received visualization service for Python code is the Jupyter Notebook software, which creates data science documents in a notebook format, however, we will skip on using this, as we will use the second option for implementation.

The second great option, which I used when implementing my data structure, and in which language I've written algorithms for my models is MATLAB (Matrix Laboratory). MATLAB is a software running its own programming language for computing and visualization written in C/C++ (which are "fast" programming languages), and a programming language, pleasant for discrete mathematics. It is a great asset for visualization, which comes in handy for presenting simulations (or the data itself) and is very commonly used by researchers. MATLAB also has implementations of graphs and graph theory algorithms, although in general more limited than Python. For example, MATLAB has a function called *maxflow* for determining the maximum flow in a graph, which is capable of finding not only the maximum flow, but can return a graph with maximum flow, and the minimum cut. However, there is no minimal cost maximum flow, or any circulation function implemented in MATLAB as of 2022. MATLAB is typically faster than Python and more comfortable for data visualization, but "has less options". Personally, I prefer MATLAB also because of the table way of storing variables. It comes handy when developers need to work with matrices.

The third option is with lower-level programming languages, such as C or C++. These are the fastest human-readable programming languages, since they directly compile to assembly code. (In contrast, the compilers compile Python code to bytecode, which then is not compiled to machine code, but rather read by an interpreter and commands are executed there, using a virtual machine.) One of the major disadvantages of these two languages is that while their commands are quite general, there's only so few of them, thus not many tools and structures are implemented in them, vastly nothing in discrete mathematics, making them a not very suitable choice for us. For example, they do not have the map data structure implemented, frequently used for adjacency lists. In addition, C is not object oriented, implementing graphs in C would be considerably more time consuming. There are C++ libraries on the internet for graphs, for example, the Boost Graph Library. This includes an adjacency-list-like data structure, and common algorithms as functions. (In fact, the Python package called graph-tool mentioned above is based on the C++ Boost Graph Library.) Another library is the Lemon Graph Library, development lead by the Eötvös Lóránt University on Budapest, with the help of developers from the Budapest University of Technology and Economics. This library is quite broad, optimal for operational research, but it's not very convenient. C++ code is harder to integrate, and more challenging to complement. On top of that, there are little to no visualization opportunities. If speed is not a very important factor and can be neglected, personally, I do not recommend using this option as development time may increase substantially, unless the developer is quite experienced with the C++ programming language. If speed is important, I would first recommend considering using the Python module graph-tool, which has its core structure written in C++ instead of Python, thus is comparably fast to C++.

We implement an adjacency-list-like structure in MATLAB (note that in MATLAB, the inputs will be stored as matrices after being read, which is great because dynamic memory allocation is not needed then).

## 4.1 Describing the models

### 4.1.1 Data structures for graphs in general

For storing directed weighted graphs in memory (coding data structures, which represent graphs), there are two main approaches: using a 2D adjacency matrix or using an adjacency list. The adjacency matrix of a vertex-labelled (directed) simple graph with vertices labelled as *1, 2, ... N* is an $N \times N$ matrix where $E_{ij}$ represents the edge from vertex *i* to *j*. This commonly is implemented with a two-dimensional array (or list, e.g., if the graph is expanding overtime, dynamic memory allocation is used), and can be visualized very simply, with matrices.

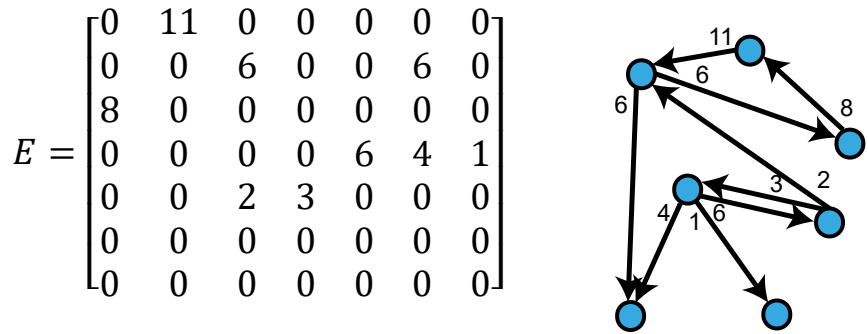

**Figure 4-1.: A graph and its adjacency matrix**

By definition, the entries in the main diagonal are zero, as simple graphs don't have edges from a vertex ending in the same vertex. This graph representation is convenient, and the lookup time for edge $E_{ij}$ is $O(1)$, constant for any edge, and we can check whether the graph contains a certain edge in constant time. However, no matter how dense the graph is, the allocated memory is always maximal. If the edges in the graph are integers, then using the *int* data type, which is, by standard, 4 bytes in size, a 100000-vertex graph requires almost 40GB of space (a bit less, since a GB is not $10^9$ bytes, but $(2^{10})^3 = 1,073,741,824$ bytes). Large networks in practice are often quite sparse, like scale-free networks, and certainly social networks, thus a large chunk of allocated memory is for non-present edges. However, for dense graphs, and especially for complete graphs, this method of representing the graph seems the best option.

For sparse graphs, a more commonly used data structure is called the adjacency list. It represents the graph as multiple lists, the *i*-th list describing the edges coming from vertex *i*. The edge lookup time in general is *O(E)*, but in simple graphs, it is at most *O(V)*, as we can look up an edge's weight by running through the list that describes the vertex's neighbours. Checking whether an edge exists in the graph in a simple graph is at most *O(V)*, as we can just go through the list of the edges coming from the starting vertex of the edge we are looking for. Financial networks that we're dealing with are quite large, however fairly sparse. This makes sense, as there is limited capacity for each participant to negotiate contracts with other participants. (Note that adjacency lists are commonly stored as a map data structure, mapping nodes to edges. This way of storing graphs with multiple properties on edges (capacity/upper limit, demand/lower limit, edge cost, possibly flow traversing the edge) seem superior, the mapping could be a list of sets, similar to this: $V_i \rightarrow [(V_j, g_{ij}, f_{ij}, k_{ij}), (V_k, \ldots), \ldots]$.

However, designing a convenient data structure to store a graph is not necessarily a goal here. In most cases, when a large data set is given (to compress), the way of storing data is quite unique (there are no universal rules for storing data structures), and the fastest way of obtaining the result we're looking for is to process the data in its regular form and stick with it, instead of developing code to transform the original data structure to our data structure design.)

### 4.1.2 Importing data into MATLAB

MATLAB is very convenient for importing well described/generated data into matrices (tables) in the MATLAB programming language. It is great for obtaining data from a .txt file or from Microsoft Excel. This comes in handy for simulations, and further usage of the algorithm. A video representation of this can be seen on YouTube: [How to Import Data from Text Files Interactively in MATLAB 2012b - YouTube](), also [How to Import Excel Data into MATLAB - YouTube](). Both are very user friendly; no code is needed to be written.

### 4.1.3 Program code of the model structure in MATLAB

In MATLAB, given the starting and ending nodes of edges (plus optionally, the weights), one can create a graph data structure with the *digraph*() command. This returns a graph object, with methods such as *Edges*, that lists the table of the edges, we can add new edges, new nodes, name the nodes, even add new variables to nodes or edges (with *G.Edges.Demands,* for example), analyse the structure of the graph, or acquire the adjacency and incidence matrices. We assign costs to the edges too, not just capacities, and each edge has some flow passing through it, so we will need to add these to the edge properties.

This is a sample code:

```
from = [1 1 2 3 3 4 5 5 6];
ends = [2 4 3 2 7 5 6 3 4];
capacities = [2 7 3 5 4 2 8 3 3];
edgeflow = [0 0 2 2 0 1 1 0 1];
demands = zeros(size(from));
costs = -ones(size(from));
EdgeTable=table([from' ends'], capacities' , edgeflow' , demands' ,costs',
'VariableNames',{'EndNodes' 'Capacity' 'Edgeflow' 'Demands' 'Costs'});
names = {'1' '2' '3' '4' '5' '6' '7' '8'}';
NodeTable = table(names,'VariableNames',{'Name' });
G = digraph(EdgeTable,NodeTable);
```

Notice the similarities with the map representation of the adjacency list. We have a list of nodes (*names* here), to which an edge table is mapped, containing the start and endpoint, capacity, demand, flow and cost of each edge. It is mapping each vertex *v* to the set of edges coming from it, edges having many properties, basically: $v \rightarrow \{\forall u\ (v, u, f_{vu}, x_{vu}, g_{vu}, k_{vu}) | vu \in A\}$.

We can also plot out the resulting graph to get a visualization of it:

```
plot(G,'NodeLabel',G.Nodes.Name,'EdgeLabel',G.Edges.Capacity);
```

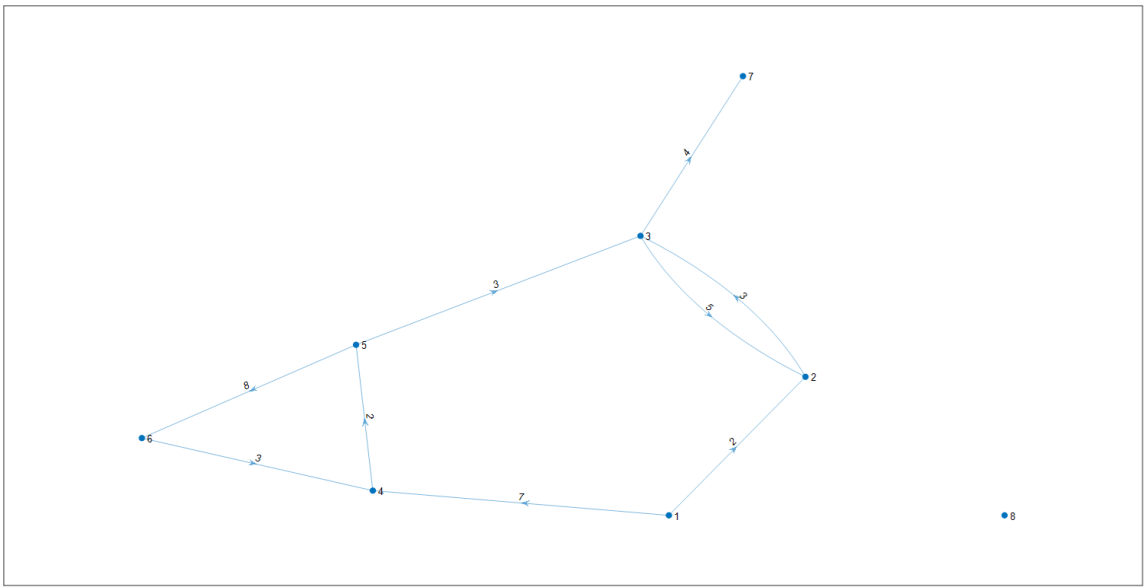

**Figure 4-2.: The plot of the graph above.**

To check whether the edgeflow function is indeed a circulation, we could use this piece of code:

```
zerotable = zeros(size(G.Nodes));
Sums = table(zerotable,'VariableNames',{'Difference' });
Nodesum = [G.Nodes Sums];
circulation = "True";
for v = 1:(size(G.Nodes,1))
    nodechecked=Nodesum.Name(v,1);
    for i=1:(size(G.Edges,1))
        startnode = G.Edges.EndNodes(i,1);
        endnode = G.Edges.EndNodes(i,2);
        if(ismember(startnode,nodechecked))
            Nodesum{v,2}=Nodesum{v,2}-G.Edges.Edgeflow(i,1);
        end
        if(ismember(endnode,nodechecked))
            Nodesum{v,2}=Nodesum{v,2}+G.Edges.Edgeflow(i,1);
        end
    end
end
```

```
    for v = 1:(size(G.Nodes,1))
        if not(Nodesum{v,2}==0)
            circulation = "False";
        end
    end
end
```

This code stores "node sums" (sum of the flow going into a node minus the flow coming out of it) in an $N \times 2$ matrix, where the first column stores each node (their name), and the second column stores the value on that node. The reason for this method is that while iterating through edges and vertices, it's very simple to compute the sum this way, as we do not have to iterate through the vertices twice this way, and it gives shorter code.

### 4.1.4 Pseudo-code of the preferential compression algorithm

For the preferential compression algorithm, a pseudo-code (size of $V$ and $A_P$ is given as $N$ and $M$):

```
import vertices as V: node list(N)
import preferred_edges as Ap: edge list(M)
import capacities as f: weights(M)
import preferences as Pref(P,fP) : list(N) of edge list (N) and list of numbers (N) #ordered list

A = Ap[0:N-1] ; #just preallocating memory
f' = f[0:N-1] ; #-||-

Gp0 : graph of V, Ap, f ;
Gp : list(M) of graph of V, Ap, f ;
Gc : graph of V, A, f';  #only N number of edges
*C : pointer to list(N) of Cs: edge list(N); #cycles

flag : number ;
i : number ;

flag = 0 ; #for edge and node deletion
i = 0 ;

#The subroutine:
```

```
subroutine "clearTTC":

    #Construct the most preferred outneighbours graph
    for j = 0:listsize(Pref)
        A[j] = Pref[j].P[0] ; #(j+1)th participant's most preferred edge
        f'[j] = Pref[j].fP[0] #and it's capacity
    end_for
    Gp[i] = digraph(V,Ap,f) ;
    Gc[i] = digraph(V,A,f') ;
    Gpi' = Gp[i]; %a copy to work on
    *C = allCycles(Gpi') ; #it returns the list of addresses in each list
#                          #of cycles, so we can later modify Gpi'
    for j = 1:listsize(*C)
        min = minimumWeight(*C[j]) ;
        for *a in C[j]
            Gpi'.f'(a) = Gpi'.f'(a) - min ;
            if(Gpi'.f'(a) == 0)
                removeEdgeFromGraph(a, Gpi') ;
                flag = 1 ; #this will surely run atleast once
            end_if
        end_for

    end_for

    while(flag ≠ 0)
        flag = 0;
        for *v in Gpi'.V
            if(Nout(v, Gpi') == ∅ or Nin(Gpi') == ∅) #set of
#                                  #outneighbours and inneighbours
                removeNodeFromGraph(v, Gpi') ;
                flag = 1;
            end_if
        end_for
    end_while

    Gc[i+1] = Gpi' ;
    i = i + 1 ;

    #If there are still nodes in the graph, run again
    if(Vi ≠ ∅)
        call subroutine "clearTTC"
    end_if

end_subroutine
```

The combination of cycles in $C$ give the circulation to compress among.

## 4.1.5 Pseudo-code and program code of the maximum volume circulation algorithm

Along with the pseudo-code, I also show a possible implementation of the maximum volume circulation algorithm's core part. Since $g = 0$ and there is at least one directed cycle in the graph, we can assure that there exists a circulation in the graph (with volume bigger than 0).

Pseudo-code:

```
Subroutine Main():
    g = new graph ;
    g = readGraphData("data.txt");
    for i=(0,size(g.Edges)-1){
        g.Edges[i].Cost = -1;
    }
    flow[size(g.Vertices)] = ∅;
    tau = -1;
    while(tau < 0){
        [flow,tau] = Min_mean_cycle(g,flow);
    }
return flow

Subroutine Min_mean_cycle(g,flow):
    size(g.Nodes) = n;
    size(g.Edges) = m;
    resg = new graph;
    resg.Nodes = g.Nodes;
    #Add the forward edges first
    for i=(0,m-1){
        if (g.Edges[i].Capacity>flow[i]){
            forward = new Edge();
            forward.Capacity = g.Edges[i].Capacity-flow[i];
            forward.StartVertex = g.Edges[i].StartVertex;
            forward.EndVertex = g.Edges[i].EndVertex;
            forward.Cost = g.Edges[i].Cost;
            resg.Edges += forward;
        }

    }
    forwardEdgesSize = m;

    #Add the backwards edges now
    for i=(0,m-1){
        if (flow[i]>0){
            backward = new Edge();
            backward.Capacity = flow[i];
            backward.StartVertex = g.Edges[i].EndVertex;
            backward.EndVertex = g.Edges[i].StartVertex;
            backward.Cost = -g.Edges[i].Cost;
            resg.Edges += backward;
        }

    }

    m = size(resg.Edges); #Update size

    #Create distance matrix and walk ending edge matrix
    d[n][n+1] = inf;
    p[n][n] = null;
    d[all][0] = 0;
    edgecosts[n][n];
    for (i,j)=(0:n-1,0:n-1){
```

```
                edgecosts[i][j] = edgeIsIn(resg,i,j)? cost(resg,i,j) : 0; #Elvis
    #                                                                    #operator
     }

    for k=(0:n-1){
        for j=(0:n-1){
            d[j][k+1] = minPairSum(d[all][k],edgecosts[all][j]);
            p[j][k]                                                     =
 edgecosts[minPairSum(d[all][k],edgecosts[all][j]).Row,j]
        }
    }

    max_diffs[n]=-inf;
    for v=(0:n-1){
        for k=(0:n-1){
            if(max_diffs[v] < (d[v][n]-d[v][k])/(n-k) )
                max_diffs[v] = (d[v][n]-d[v][k])/(n-k)
        }
    }
    (min_mean,min_mean_vertex) = (min(max_diffs),min(max_diffs).index)

    #Find cycle d[v][n]
    d_nWalk = walk(p[v][all]);
    C = cycles(d_nWalk);
    min_cycle = null;
    for (c in C){
        if(costsum(c)==min_mean){
            min_cycle = c;
            continue
        }
    }
    min = minimumCapacity(min_cycle);
    for j=(1:m){
        if(edgeIsIn(min_cycle,resg.Edges[j])){
            flow[j < forwardEdgesSize ? j : j-forwardEdgesSize] +=
 resg.Edges[j].Cost*tau
        }

    }
 return flow, tau
```

The minimum mean cycle (assuming that all demands are 0) algorithm is implemented as a function. The reason to this is that we look for the minimum mean cycle multiple times, not just once (but at most $4nm^2 \lceil \ln(n) \rceil$ times, as shown by Goldberg and Tarjan).

The function takes the endpoints of edges, their capacities, costs, and flow amount, and returns a circulation improved among the minimum mean cycle. I first create variables for the residual graph: forward and reverse edges described as their endpoints, capacities and costs.

```
function [circ_new,tau]=
min_mean_cycle_0demand(from,ends,capacities,costs,circ_current)

%Circulation has demand function g=0 for simpilicty
demand=0;
%Create the residual graph
forward_from=[];
forward_ends=[];
forward_capacities=[];
forward_costs=[];
reverse_from=[];
reverse_ends=[];
reverse_capacities=[];
reverse_costs=[];
```

Since $g = 0$ and there is at least one directed cycle in the graph, we can assure that there exists a circulation in the graph (volume can be 0). We fill up the row vectors that represent the residual graph with the given data:

```
for i = 1:size(circ_currect,2) %loop through the circulation
   if(circ_current(1,i)<capacities(1,i))
        forward_from = [forward_from from(1,i)];
        forward_ends = [forward_ends ends(1,i)];
         forward_capacities  =  [forward_capacities  ( capacities(1,i)-
circ_current(1,i) )];
        forward_costs = [forward_costs costs(1,i)];
    end
   
   if(circ_current(1,i)>0)
        reverse_from = [reverse_from ends(1,i)];
        reverse_ends = [reverse_ends from(1,i)];
        reverse_capacities = [reverse_capacities circ_current(1,i)];
        reverse_costs = [reverse_costs costs(1,i)*(-1)];
    end
end
res_from = [forward_from reverse_ends];
res_ends = [forward_ends reverse_from];
res_capacities = [forward_capacities reverse_capacities];
res_costs = [forward_costs reverse_costs];
resG = digraph(res_from, res_ends, res_capacities);
resG.Nodes=(1:n)';
resG.Edges.Costs=res_costs;
n=size(resG.Nodes,1);
m=size(resG.Edges,1);
```

Converting it to an internal data structure with digraph is convenient, but not necessary.

Now, let's create the distance matrix (containing $d_k(v_i)$ for $0 \leq k \leq n, 1 \leq i \leq n$), and fill it up with the corresponding values. We also store the edges at the end of walks setting the values in the distance matrix.

```
%Creating an Nx(N+1) matrix for distances, filling it up with Inf values
```

```
distance_matrix = ones(size(resG.Nodes,1),size(resG.Nodes,1)+1)/0;
distance_matrix(:,1)=0; %Fill up first column with 0 (d_0(u)=0)
settingedge_matrix = -ones(n)/0; %negative infinities
%Finding distances (similar to Bellman-Ford)
for k=1:n
    for j=1:m
        if(distance_matrix(res_from(1,j),k)+res_capacities(1,j)*res_costs(1,j)<distance_matrix(res_ends,k+1))

distance_matrix(res_ends,k+1)=distance_matrix(res_from(1,j),k)+res_capacities(1,j)*res_costs(1,j);
        end
    end
end
```

Now we search for the minimum mean cycle. First, we check the $\frac{d_n(u)-d_k(u)}{n-k}$ values for each $u \in V$ and $0 \leq k \leq n$. For each $u \in V$, we store the maximum value too.

```
%Find the minimum mean cycle
maxdiffs = -ones(1,n)/0; %negative infinities
for v=1:n
    for k=1:n
        if(((distance_matrix(v,n+1)-distance_matrix(v,k))/(n+1-k))>maxdiffs(1,v))
                                      maxdiffs(1,v)=(distance_matrix(v,n+1)-distance_matrix(v,k))/(n+1-k);
        end
    end
end
```

Here, we check the minimum of these values, and the vertex that sets the minimum. This gives us the minimum mean of a cycle. (Assume that vertex is $v$.)

```
min_mean = maxdiffs(1,1);
for v=1:n
    if (min_mean>maxdiffs(1,v))
        min_mean=maxdiffs(1,v);
    end
end
```

If this value is at least 0, then the circulation already has maximum volume, else we can improve the circulation. Thus, we either return (see the end of the code), or we improve the circulation.

```
%If the minimum mean is less than 0, then we can improve on the cost, else
%we are finished
if(min_mean<0)
    %Find the minimum mean cycle
```

A minimum mean cycle is contained in $d_n(v)$, so we construct the walk that set the value of $d_n(v)$ from the edges stored above, knowing that $d_{k+1}(v) = \min\{d_k(u) + c(a) | a = (u,v) \in A\}$. Store the vertices only.

```
%Store the vertices in dn(v)y
settingwalk = 1:(n+1);
for k = 1:n
    settingwalk(r) = res_from(settingedge_matrix(min_mean_vertex,k));
end
settingwalk(n+1) = res_ends(settingedge_matrix(min_mean_vertex,n+1));
```

We now allocate memory for the minimum mean cycle, and also for its ending vertex (that's also the starting vertex).

```
min_mean_cycle = 1:n;
min_mean_cycle_end = n;
```

The method I implemented for finding such a cycle is by iterating through nodes of the walk once, and while iterating, simultaneously iterate once more at every iteration (with another variable). If the two iterations at the same time result in the same node, then we have a cycle, and we compute its mean cost.

```
for v = (2:n+1)
    for k = (1:v-1)
        if(settingwalk(k)==settingwalk(v))
            cycle = 0;
            for c = (k+1:v)
                cycle = cycle + res_costs(settingedge_matrix(min_mean_vertex,c));
            end
```

If this cycle's mean cost is equal to the minimum mean, we've found a minimum mean cost cycle.

```
            if(cycle/(v-k) == min_mean)
                for c = (k+1:v)
                    min_mean_cycle(c-k) = settingedge_matrix(min_mean_vertex,c);
                end
                min_mean_cycle_end = v-k;
            end
        end
    end
end
```

Now, we can evaluate the minimum residual capacity in the cycle, that will be the maximum amount to clear with.

```
%Find tau
```

```
        tau = 0;
        for k = (1:min_mean_cycle_end)
            diff_k = res_capacities(settingedge_matrix(min_mean_vertex,k));
            if(diff_k<tau)
                tau = diff_k;
            end
        end
```

At last, we update the circulation.

```
        circ_new=circ_current;
        for  k = (1:min_mean_cycle_end)
           edge = settingedge_matrix(min_mean_vertex,k);
           if(edge>size(forward_costs))
               circ_new(1,edge)= circ_current(1,edge)- tau;
           else
               circ_new(1,edge)= circ_current(1,edge)+ tau;
           end
        end

    else
        circ_new=circ_current;
        return
    end
```

From here, one can write a "main" file that gathers the data from the database, redesigns the data structure if needed, and creates a *while* loop that runs until the returned value *tau* is nonnegative, starting from an all-zero circulation and equalling it to the returned circulation in each iteration.

# 5 Possible future research and summary

This chapter concludes all the work above, and extends it further with ideas, and suggestions on improving the model and results.

## 5.1.1 Simulation of the models

To show empirical results of how good the models are in terms of fraction of excess compressed, we should run the models on artificially created graphs to gather some insight on how well the algorithm perform. (Compressing real data is great as well, but there aren't many datasets to work on, and in this case, simulating compression once on many different graphs may be more expressive, as we need a large sample size for averaging out results.)

I recommend a method for generating graphs for this simulation: Assume the graph we'd like to generate has *n* number of nodes and *m* number of arcs. To make sure there is at least one cycle in the graph, randomly generate one arc from every vertex (this is analogue to each vertex picking a most preferred other vertex). We then check how many different strongly connected components are there and keep on reducing the number of components with 2 arcs (by adding one random arc between two components in one direction and one in the other direction) until there is only one strongly connected component, then randomly generate the remaining $m - n - k$ number of new arcs ($k$ being the number of arcs added during connecting components). In later research, I will conduct this simulation.

## 5.1.2 Participants describe lower bounds for the amount of cleared notional on each of their affected contracts

On 3.2, we described the definition of circulations, some common properties of circulations, we even stated Hoffman's theorem, which states when there exists a circulation in a graph, assuming for each arc there is a lower and upper bound on how much flow can pass through the arc. In our model, we assumed that $d = 0$, thus the lower bound is zero, we only have an upper limit on the traversing flow, for convenience. In reality, assuming that each arc has a lower bound for traversing flow might be a better approach. Take for example a large network, where "at the top" of the hierarchy list are huge international banks, and at the bottom are small local businesses. A cycle large enough to contain both might not be eligible for clearing, because the clearing amount of each participant is low (even though they sum up to a large value) due to small businesses only lending small amounts, and the banks at the top may not want to join the clearing procedure as it is a lot of paper work for them, that in the end might result in more financial loss for them than gain from compressing. (Somebody needs to be paid to coordinate all that…)

If an algorithm suggests clearing among this cycle, we can reduce the obligations in this cycle by at most the amount of the lowest obligation. If the cycle is large enough, it may contain a large financial agent (an international bank e.g.) and a relatively small participant (e.g., a local business or a start-up). The small participant may only take small loans and lend even less, whilst the large company may only want to compress portfolio when the amount of their compressed excess is a considerable amount for them. Thus, in theory, it may happen, that the large financial agent rejects the offer for financial clearing in this case. To tackle this problem, we must consider the minimum excess each participant requires to excess for accepting the clearance offer. Logically, registering a lower limit for the amount of flow to pass through each node seems the most applicable, as each participant is asked to report the minimal amount of excess in their loans and obligations that they are willing to eliminate. One may approach solving this type of problems with linear programming, however, Schrijver [2] shows a graph theoretical method to extend flow, circulation or transhipment problems directly to a problem where vertices also have a lower (and/or upper) bound on traversing that vertex.

Another sufficient approach that is sufficient is if participants gave lower bounds of traversing flow on each arc, since this in itself gives the minimal amount of excess to be reduced for each participant but is significantly stricter. It is also questionable whether this is realistic in practice. Circulations with lower and upper bounds have been studied as we've shown before, so applying the knowledge from it in this context may be a solution.

Our models used algorithms that work assuming a lower limit (so $d$ could be larger than 0), however, with this exact approach, we may get heavily worse results. This is not only due to the fact that as the lower limit is increased on each arc (from 0 to a positive number), an *f* circulation that satisfies $f(a) \leq c(a)$ may not satisfy $d(a) \leq f(a) \leq c(a)$ if $d > 0$, thus *f* may not be a feasible circulation, and adjustments are needed to be made, but there may not even exist a feasible circulation, even though we surely could clear some excess amongst cycles. The reason for this method failing to give an effective clearing when there is a non-zero lower limit is that these algorithms (mainly the minimal cost circulation finding one) look for a circulation that satisfies the minimum amount of flow passing through every arc, which obviously may not necessarily exist. Large graphs can easily contain a similar small subgraph in them, where the minimum amount of flow needed to flow into the subgraph is more than the maximum amount of flow limited to flow out of the subgraph. Better yet, if we don't just look at dealers of the market but try to run the algorithm on the whole market (including customers), we will surely not have any existing circulation, as some vertices will either have arcs only coming out of them, or arcs only coming into them. Thus, for us, if we want to work with demand $d > 0$, to find the maximum amount of excess in conservative clearing, it only makes sense to not only evaluate the minimum cost circulation in the original graph (with $k = -1$), but to evaluate the minimum cost circulation in subgraphs as well. Equivalently, you could extend the feasibility on each arc by redefining a flow to be feasible if for any arc $a \in A$ either $g(a) \leq z(a) \leq f(a)$ or $z(a) = 0$. One can elaborate more on this; however, this leads to an NP-hard problem.

## 5.2 Compressing among circulations with preferences: combining the two previous models

As we have seen, the circulation function is a strong tool that can help designing a network compression and can outperform compression with multiple cycle clearing. We also gave a model for compression, where participant preferences are taken into consideration, which might to be the better model for satisfying the participants' (untold) needs. (Although one may be sceptical of the model, if clearing an obligation is very beneficial for one contracting participant (clearing the obligation they'd like to clear most must bring some advantage to them over clearing other obligations), then it most often is not beneficial for the other participant, as likely it could have gathered more profit from not letting this obligation be cleared, and if a participant calculates that they would lose more than gain, they surely reject the compression offer. This is partly the reason why I do not recommend defining preferences on which obligation companies want to be paid back to them.) Participant preferential models and system optimized models make a strong juxtaposition. It would be useful to create a model that inherits from both and satisfies participants whilst maximizing benefits for the coordinating authority. For example, one may create an algorithm that compresses maximum volume excess (via finding a maximum volume circulation) and uses preferences as a parameter.

We can gather some thought from Fleiner's work [26]. Fleiner introduced preferences to the flow problem, and defined "stable" flows, flows that do not let any blocking walks appear in the graph, i.e., a directed walk on unsaturated arcs (arcs that have room for more flow to be sent through them) such that both ends of the walk is either a terminal or can improve its position by moving some flow from a less preferred arc onto the walk. This was built on Gale's and Shapley's work [19] done on stable matchings and Baïou's and Balinski's work [27] on stable allocations. He formulates the stable flow problem and reduces it to the stable allocation problem, pointing out that in fact, the reduction resembles to the reduction of the maximum flow problem to the maximum b-matching problem. He proves the stable flow theorem, generalization of the Gale-Shapley theorem: a stable flow always exists in any graph. Based on his work, one might formulate a similar problem on circulations instead of flows. It might be true, that a circulation might exist in a graph (where demands equal to 0, or where a feasible circulation exists) with no blocking circuits. This would allow us to search for a "stable circulation", a circulation with minimum cost and most optimal in terms of preferences.

## 5.3 Summary

In this paper, upon describing the problem and our toolset, I summed up part of the literature in this topic and associated topics. I then gave a preferential model for compressing with preferences and gave a model for maximum volume conservative compression.